\documentclass[prl,twocolumn,,amsmath,amssymb,floatfix]{revtex4}
\usepackage{graphicx}
\usepackage{bm}

\newcommand{\be}{\begin{equation}}
\newcommand{\ee}{\end{equation}}

\def \be{\begin{equation}}
\def \ee{\end{equation}}
\def \ba{\begin{array}}
\def \ea{\end{array}}
\def \beq{\begin{eqnarray}}
\def \eeq{\end{eqnarray}}

\begin{document}
\title{Spectroscopy of collective excitations in interacting low-dimensional many-body systems using quench
dynamics}

\author{Vladimir Gritsev$^1$,  Eugene Demler$^1$,  Mikhail Lukin$^1$, Anatoli Polkovnikov$^2$}
\affiliation {$^1$Department of Physics, Harvard University, Cambridge, MA 02138\\
$^2$ Department of Physics, Boston University, Boston, MA 02215}

\begin{abstract}
We study the problem of rapid change of the interaction parameter
(quench) in many-body low-dimensional system. It is shown that,
measuring correlation functions after the quench the information
about a spectrum of collective excitations in a system can be
obtained. This observation is supported by analysis of several
integrable models and we argue that it is valid for non-integrable
models as well. Our conclusions are supplemented by performing exact
numerical simulations on finite systems. We propose that measuring
power spectrum in dynamically split 1D Bose-Einsten condensate into
two coupled condensates can be used as experimental test of our
predictions.
\end{abstract}

\date{\today}

\maketitle


In many cases interesting quantum phenomena appear not in properties
of the ground states but in the coherent dynamics of the system away
from equilibrium. Canonical examples from basic quantum mechanics
include Rabi oscillations in two level systems, collapse and revival
in the Jaynes-Cummings model, Landau-Zener tunneling.  Analysis of
non-equilibrium coherent dynamics in a strongly correlated many-body
systems is a more challenging task than in a single particle quantum
mechanics, hence progress has been made only in a few cases. This
includes changing parameters in the transverse field Ising
model~\cite{CL}, mean field dynamics in systems with BCS pairing of
fermions~\cite{Y}, collapse and revival phenomena~\cite{GB} and
non-equilibrium superfluid phase oscillations~\cite{topk} of bosonic
atoms in optical lattices. In this paper we discuss another example
of non-equilibrium coherent dynamics in strongly correlated many
body system. We consider a sudden quench of interaction parameter in
nonlinear interacting system and show that the subsequent time
dynamics exhibit oscillations with frequencies given by the poles of
the {\it scattering matrix} of this many-body system. The sudden
quench can be thus used as a tool for spectroscopy of the
interacting models. We make our analysis by increasing complexity of
examples: we start with a quantum Josephson junction (qJJ) model
then generalize it for a Gaussian model and support the main
observation on the the quantum sine-Gordon model (qSG). At the end
we generalize the main statement for non-integrable models and make
supporting arguments in favor of our conclusion for general models.

The qSG model, is a popular prototypical example of nonlinear
interacting quantum system. This model appears as an effective
description of variety of condensed-matter, statistical physics and
field theory problems. Thus in condensed matter it describes
low-dimensional spin and charge systems, disordered systems (see
\cite{Giam} for review) and coupled Bose condensates \cite{long}.
Given the large freedom of tunability of parameters in ultracold
quantum gases we have in mind the test of a theory presented below
in these dynamically decoupled 1D condensates.

The dynamics in qSG model can be analyzed using its exact solution.
In general, describing dynamical properties of a many-body system
using the exact Bethe Ansatz solution is not a straightforward
procedure. In our case, progress can be made connecting the problem
of time evolution from a certain initial state to the equilibrium
sine-Gordon model with a boundary. For conformally invariant models
a similar approach of mapping temporal evolution after a quench to a
class of boundary phenomena was discussed in Ref.~[\onlinecite{cc}].
In our case the system can have different mass scales corresponding
to solitons and their bound states.

The Hamiltonian of the qSG model is given by
\be {\cal H}_{SG}=\frac{1}{2}\int dx
[\Pi^{2}(x)+(\partial_{x}\phi)^{2}-4\Delta\cos(\beta \phi)].
\label{sgham}
\ee
In application to split condensates the $\phi(x)=\phi_{1}-\phi_{2}$
is the relative phase between the two of them and $\Pi(x)$ is the
conjugate momentum proportional to the difference between local
densities and $\beta=\sqrt{2\pi/K}$ \cite{long}. The interference
experiments, such as reported in Ref.~[\onlinecite{Sch}], measure
the $\phi(x)$ between the two condensates.
After the splitting the system is in a state which is not an
eigenstate, hence, this initial state will undergo a complicated
quantum dynamics controlled by the many-body Hamiltonian. We assume
that at $t=0$ the system is prepared in a state with $\phi=0$ for
all $x$. In reality $\phi$ is a wave packet with the width
determined either by the rate with which the condensates were
separated~\cite{anton} or by the depletion~\cite{Ehud}.
The Luttinger parameter $K$ of individual condensates in Eq.
(\ref{sgham}) is large for weakly interacting bosons, and approaches
1 in the hard core repulsion (Tonks-Girardeau) regime. $K\leq1$ for
fermionic and spin systems. Generally $K$ can be extracted from the
microscopics \cite{Giam}.

We concentrate our analysis here on dynamical properties of
one-point correlation function, however generalization of our
approach for the multi-point correlation functions is
straightforward. The one-point correlation function $\int_{0}^{L}
dx\langle \psi(t=0)|\exp(i\beta\phi(x,t)|\psi(t=0)\rangle$
corresponds to the
amplitude of interference fringes, $A(t)$\cite{Sch} and can be
measured as a function of the evolution time $t$.
To characterize the time
evolution of $A(t)$ we analyze the power spectrum $P(\omega)=
\lim_{T\rightarrow \infty} |\int_0^T dt\, e^{i\omega t} A(t) |^{2}$.

It is instructive to analyze first a simpler situation, when the
spatial fluctuations of the phase $\phi$ are energetically forbidden
and the Hamiltonian (\ref{sgham}) reduces to the qJJ model
\begin{eqnarray}
{\cal H}_{JJ}=\frac{E_c}{2}\left(\, \frac{\partial}{i\,
\partial\phi} \,\right)^{2}-J\cos(\phi)
\label{QJJ}
\end{eqnarray}
with $E_c=\beta^{2}\hbar v_{s}/L$ and $J=2L\Delta$ where $v_{s}$ is
a sound velocity. Following the standard approach for analyzing
sudden perturbations in quantum mechanics we decompose the initial
state into eigenstates of the Hamiltonian (\ref{QJJ}), $|\psi (t=0)
\rangle = \sum_n a_n | n \rangle$. Eigenstates $ | n \rangle$ can be
found explicitly using Matthieu's functions\cite{RG}. Only even
$n$'s are present in decomposition of $|\psi (t=0) \rangle$ and the
occupation probabilities $ |a_n|^2$ decrease with increasing $n$,
provided that the initial state has a finite width. Letting system
to evolve for time $t$ we find a state $|\psi (t) \rangle = \sum_n
a_n e^{-i \omega_n t} | n \rangle$. The amplitude $A(t) = \langle
\Psi(t) | \cos(\phi) | \Psi(t) \rangle = \sum_{nm} a_n^* a_m ( \cos
\phi )_{nm} e^{-i(\omega_n-\omega_m)t}$, where $( \cos \phi)_{nm}$
is the matrix element of $\cos \phi$ between the states $n$ and $m$.
The power spectrum for a specific set of parameters is shown in
Fig.~\ref{fig_jospehson}. The inset on this graph shows the energy
level structure and a couple of possible oscillation frequencies.
\begin{figure}[ht]
\includegraphics[scale=0.80,angle=0,bb=00 10 250 210]{quench_josephson1.eps}
\vspace{-0.5cm} \caption{Power spectrum for a single qJJ consisting
of two sites. At $t=0$ the tunneling amplitude is suddenly reduced
from a very large value to $J=10$, the interaction strength $E_c=1$.
The inset shows the energy levels of the qJJ with the arrows
indicating a couple of possible oscillation frequencies. }
\label{fig_jospehson}
\end{figure}
The power spectrum consists of different peaks corresponding to the
overlaps between various even energy levels. The largest (central)
peak in the power spectrum comes from beating of $n=0$ and $n=2$
states (i.e. the $(\cos \phi)_{20} e^{i (\omega_2-\omega_0)t} + {\rm
c.c.}$ term). The dominance of this peak comes from a combination of
the two factors: i) The matrix elements $(\cos \phi)_{nm}$ are the
largest for the smallest difference between $n$ and $m$ ii) The
state $n=0$ has the largest weight in the decomposition of $|\psi
(t=0) \rangle$. There are also contributions at frequencies
$\omega_m-\omega_0$ for $m=4,6,...$ (only $m=4$ is shown), which
give rise to peaks at higher frequencies. For a harmonic potential
such peaks would appear at precise multiples of the frequency of the
central peak. But because of the unharmonicity such peaks are
shifted to lower frequencies. Similarly other important
contributions to the power spectrum come from beating between states
$n$ and $n+2$ with $n>0$. They appear at smaller frequencies than
that of the central peak. Finally we have contributions from
beatings of various other combinations of states, but they come with
a smaller weight.

We now discuss a case of full 1D model (\ref{sgham}). In the large
$K$ limit, the qSG model can be well approximated by the Gaussian
theory with a massive term $\sim m\phi^{2}$, where $m\sim
J_{\perp}$, instead of the cosine. The initial state is then a
product of {\it squeezed} states for all $k$ vectors
$|\psi_0\rangle\sim \prod_k \exp(a_k^\dagger
a_{-k}^\dagger)|0\rangle$, where $a_k^\dagger$ is the usual bosonic
creation operator for the excitation with the momentum $k$ and the
energy $\omega_k=\sqrt{k^2+m^2}$. Expanding it one observes that the
state $|\psi_0\rangle$ contains various combinations of pairs of
particles with opposite momenta.
The leading contribution to $A(t)$ coming from two-particle
excitations behaves at long times as $\langle
\cos(\phi(x,t))\rangle\sim \sin(2 m t)/\sqrt{t}$. This behavior
leads to the power law singularity in the power spectrum:
$P(\omega)\sim |\omega-2m|^{-1}$. Higher harmonics corresponding to
multi particle excitations correspond to weaker singularities and
their weights are suppressed.

With decreasing $K$ nonlinearities start to play increasingly
important role. As we will show below the peak at fundamental
frequency $\omega_0$ splits into a sharp singularity and a
two-particle contribution (see Fig.~\ref{sg-ps}). The two-particle
contribution corresponds to the excitation of a pair of two lowest
energy breathers $B_1$, which are direct analogues of the massive
excitations in the Gaussian model. The singularity corresponds to
the non-decaying excitation of an isolated $B_2$ breather, which is
a bound state of two $B_1$ breathers (for more explicit notations
see below). In addition to splitting of the fundamental peak one
finds that higher harmonics are shifted to lower frequencies in a
direct analogy with a qJJ. Beatings of different harmonics lead to
appearance of singular peaks at frequencies smaller than $\omega_0$,
similarly to a qJJ system.

For finite $K$ a convenient complete basis of the qSG model is
provided by the asymptotic scattering states which can be obtained
by the action of the elements $A_{a_{k}}(\theta)$ of the
Zamolodchikov-Faddeev algebra~\cite{Smirnov} on the vacuum state:
$|\theta_{1}\theta_{2}...\theta_{n}\rangle_{a_{1},a_{2},...,a_{n}}=
A^{\dag}_{a_{1}}(\theta_{1})A^{\dag}_{a_{2}}(\theta_{2})...A^{\dag}_{a_{n}}(\theta_{n})|0\rangle$.
Here the operators $A_{a_{k}}(\theta_{k})$ have internal index
$a_{k}$ (corresponding to solitons (+), antisolitons (-) or
breathers (n)(denoted by $B_{n}$ below)) and depend on the rapidity
variable $\theta_{k}$, defining the momentum and the energy of a
single quasi-particle excitation of mass $M_{a}$:
$E=M_{a}\cosh(\theta)$, $P=M_{a} \sinh(\theta)$. The operators are
defined in such a way that $A_{a_{k}}(\theta_{k})|0\rangle =0$,
$\forall k, a_{k}$. Explicit dependence of $M_{a}(\Delta)$ was found
in Ref.~[\onlinecite{Zmass}].

Next we consider the initial condition $\phi(t=0)=0$ as a {\it
boundary-in-time} condition of the Dirichlet type. It belongs to the
class of integrable initial (boundary) conditions, which generically
have the following form~\cite{GZ}:
\beq\label{Bstate} |B\rangle ={\cal
N}e^{\sum_{a,b,n}(\frac{g_{2n}}{2}B_{2n}^{\dag}(0)+\!\int\limits_{-\infty}^{\infty}
\frac{d\theta}{4\pi}K^{ab}(\theta)A^{\dag}_{a}(-\theta)A^{\dag}_{b}(\theta))}|0\rangle.
\eeq
The matrix $K$ is related to the reflection matrix of solitons,
antisolitons and breathers. These matrices are known~\cite{GZ} for
all integrable boundary conditions. The important feature of these
boundary states is the presence of various poles in the reflection
amplitudes $K^{ab}$. These poles correspond to bound states of the
particles $A_{a}, A_{b}$.  In the soliton-antisoliton channel such
bound states are breathers with masses
$M_{B_{n}}=2M_{s}\sin(n\pi/(8K-2))$, where $n=1,\ldots,4K-1$ and
$M_{s}\equiv M_{s}(\Delta,K)$\cite{Zmass}.
\begin{figure}[ht]
\includegraphics[scale=0.8,angle=0,bb=00 30 250 210]{ps_K=1_6_d=0_4.eps}
\caption{Power spectrum for $K=1.6$ and $\Delta=0.4$ including
single and two-breather contributions. Many other contributions are
not visible on this scale. Arrows indicate $\delta$-peaks of type
(i) (see the text below). The inset shows the power spectrum from
the exact diagonalization on $6\times 2$ sites for the bosonic
Hubbard model. Note the break in the vertical scale. Frequency units
are defined by $M_{s}(\Delta,K)$ for given $\Delta$ and $K$.}
\label{sg-ps}
\end{figure}
The poles in the soliton-antisoliton reflection amplitude $K^{+-}$
imply the possibility of exchanging breathers with zero momentum,
$B_{a}(0)$. The corresponding $``+-B_{a}"$ coupling is denoted by
$g_{a}$. For more discussions and results on the structure of
boundary states see \cite{BPT}. We note that the state
(\ref{Bstate}) is a generalization of squeezed states in the
Gaussian theory. Presence of additional breather terms at zero
rapidity is the consequence of unharmonicity of the underlying
model. Another important difference with the harmonic theory are the
unusual commutation relations between $A_{a}$'s which are neither
bosons nor fermions. The state (\ref{Bstate}) is the initial state
in our problem.  The Hamiltonian (\ref{sgham}) is diagonal in $A$'s,
which allows us to compute the time dependence of $|\psi_{B}\rangle$
\beq\label{time}
&&|\psi(t)_{B}\rangle = {\cal
N}\exp\biggl[\sum_{a,b,n}\frac{g_{2n}}{2}B_{2n}^{\dag}(0)e^{-iM_{B_{2n}}t}\\
&&+\int_{-\infty}^{\infty}\frac{d\theta}{4\pi}\exp(-iMt\cosh(\theta))
K^{ab}(\theta)A^{\dag}_{a}(-\theta)A^{\dag}_{b}(\theta)\biggr]|0\rangle\nonumber
\eeq
where $M=M_{a}+M_{b}$.
For a translation invariant system $A(t)= L \langle \psi(t) | e^{ i
\beta\phi (x=0)} | \psi(t) \rangle$ where $L$ is a system size.
We evaluate $A(t)$ by expanding the exponential form of
$|\psi(t)_{B}\rangle$ and computing terms one by one.  We use the
form-factors (FF) technique discussed by us in
Ref.~[\onlinecite{long}] for a related problem (for a general review
of the FF approach see Ref.~[\onlinecite{Smirnov}]). Our
calculations rely on recent progress in evaluation of FF for the qSG
model~\cite{Luk}. In general, FF are defined as expectation values
of the operators in the asymptotic states
$|\theta_{1}\ldots\theta_{n}\rangle$, which can always be
represented in the canonical form, $F^{{\cal O}}=\langle 0|{\cal
O}|\theta_{1}\ldots\theta_{n}\rangle_{a_{1}\ldots a_{n}}$. It is
known from general principles \cite{Smirnov} that the FF expansion
converges rapidly with increasing number of the participating
states. We will therefore discuss the most important contributions
to the power spectrum, which come from the first terms in the
expansion of Eq.~(\ref{Bstate}). These contributions split into the
following categories: i). Delta function peaks corresponding to
contributions of individual breathers. They appear at frequencies
$\omega^{(1)}=M_{B_{2m}}$  and
$\omega^{(11)}=M_{B_{2m}}-M_{B_{2n}}$, $m\geq n$. The strengths of
these peaks are given by $g_{m}F_{B_{2m}}/2$ and
$g_{m}g_{n}F_{B_{2m}}F_{B_{2n}}/4$ correspondingly. Here the
single-particle breather FF's are~\cite{long}:
\be\label{1B} F_{B_{n}}^{\exp(i\beta\phi)}\!=\!\frac{{\cal
G}_{\beta}\sqrt{2}\sin(\pi n\xi)\exp[I(-\theta_{n})]e^{\frac{i\pi
n}{2}}}{\tan(\frac{\pi\xi}{2})(\cot(\frac{\pi\xi
n}{2})\prod_{s=1}^{n-1}\cot^{2}(\frac{\pi \xi s}{2}))^{1/2}}, \ee
where $\theta_{n}=i\pi(1-n\xi)$, $\xi=1/(4K-1)$, $I(\theta)$ is
given in \cite{long}
and the numerical factor ${\cal G}_{\beta}$ was computed in
Ref.~[\onlinecite{LZg}]. These contributions generally decrease with
$n$ and decrease with $K$. ii) Two-particle $A_{a}A_{b}-0$
contributions of excitations with equal masses corresponding to
$\omega^{(2)}=2M_{A_{a}}$. They come from a continuum part. Their
weights can be estimated by multiplying corresponding single
particle FF's. iii) Interference contributions involving more than
two particles of equal or unequal masses.

In Fig.~\ref{sg-ps} we present the results of our calculations for
specific values of $K$ and $\Delta$. For large $K$ the dominant
contribution comes from beating of the vacuum state and the
$B_1(\theta)B_1(-\theta)$ pairs corresponding to the massive
phonons. The corresponding contribution is absent for a QJJ. The
reason is that $B_1$ breathers can be excited only with nonzero
momentum. However, the non-decaying peaks, corresponding to
particles excited at strictly zero momentum, have their direct
analogues (see Fig.~\ref{fig_jospehson}). Indeed the $B_2-0$ peak
corresponds to the ``central peak"  in the QJJ picture. The
satellite peaks $B_4-B_2$, $B_6-B_4$, $B_4-0$ etc. also have their
analogues in Fig.~\ref{fig_jospehson}. Such peaks correspond to the
non-decaying oscillations. Their $\delta$-peak nature is a
consequence of the qSG integrability.

Our analytic results are supported by exact numerical simulations in
a system of $N=12$ particles in two chains each of 6 sites. We used
the Hubbard model with periodic boundary conditions and with the
interaction $U=1$ and intrachain hopping $J=1$ to realize a system
with moderate value of $K$ far from the Mott state . At time $t=0$
the inter-chain hopping $J_\perp$ was abruptly decreased from a very
large value to $J_\perp=0.1$. Despite seemingly small size we point
that this system contains more than $10^6$ states, which is more
than enough to distinguish integrable versus nonintegrable dynamics
(see e.g. Ref.~[\onlinecite{Vadim}]). The results of such
simulations are in complete qualitative agreement with the analytic
predictions (see the inset in Fig.~\ref{sg-ps}). We can identify
peaks corresponding to various breathers. There are also two
additional two-particle contributions denoted as $B_1B_1-0$ and
$B_2B_2-0$ in the inset. These peaks correspond to excitations with
nonzero relative momentum. However unlike in the true thermodynamic
limit, these oscillations are not broadened, since there is no
continuum of different momentum states.

Previous analysis for the integrable models can be generalized
further for non-integrable models using the form-factor perturbation
theory \cite{DMS}. If we consider not-so-strong deviations from
integrability, which means that there are still no multiple particle
production in a theory (which is equivalent to the absence of a
branch-cuts in a physical strip of a Mandelstam variable), one can
argue that the first effect is the change of the position of the
poles in the scattering matrix. This variation of pole can be shown
to be given by the particle-antiparticle FF
$F(\theta_{1}-\theta_{2})$, and the change of the particles mass is
given by \cite{DMS} $\delta m_{a}^{2}\sim F_{a\bar{a}}^{{\cal
O}}(i\pi)$ corresponding to perturbing operator ${\cal O}$.
Therefore the leading effect of deviation from integrability is a
shift of positions of peaks.

It is possible to argue in general that the time evolution of
expectation value of some local observable operator in initial state
formed by a sudden quench will provides the information about the
spectrum of the theory after this transition. One can indeed show
\cite{BBT} that the boundary reflection amplitude $R(k)$ is related
to the bulk Green function $G(x,x';t-t')$ as $G(x,x';t-t')=\int
d\omega\frac{e^{-i\omega(t-t')}}{4\pi
k(\omega)}(e^{ik(\omega)|x-x'|} +R(k)e^{ik(\omega)|x-x'|}) $ for the
field theory with dispersion $k(\omega)$. This formula is somehow
reminiscent to the $T$-matrix formulation of the impurity scattering
problem. We therefore observe that the poles structure of the Green
function is encoded in a boundary reflection factor. Another part of
the boundary state is the form-factor. The form-factors expectation
values can be expanded using the LSZ
formula\cite{LSZ} which relates the $n$-particle form-factors of
operator ${\cal O}$ to a $n$-point function. All that leads to
conclusion that spectrum generation during the quench dynamics is a
generic phenomena.


In this paper we showed that the dynamics after quench in many-body
interacting system can be used for spectroscopy of collective
excitations. We have explicitly demonstrated this on quantum
sine-Gordon type models and argued that this conclusion is valid for
more general, non-integrable models. We supported our statement for
general many-body systems using connection of the boundary
reflection amplitude to the Green's functions.

We would like to acknowledge E. Altman, I. Bloch,  R. Cherng, C.
Kollath, S. Lukyanov, M. Oberthaler, J. Schmiedmayer, G. Takacs, V.
Vuletic for useful discussions. This work was partially supported by
the NSF, Harvard-MIT CUA  and AFOSR. V.G. is also supported by the
Swiss NSF, A.P. is supported by AFOSR YIP.

\end{document}